\documentstyle[twocolumn,amssymb,aps,epsfig,float]{revtex}
\begin{document}
\bibliographystyle{plain}
\twocolumn[\hsize\textwidth\columnwidth\hsize\csname @twocolumnfalse\endcsname

\title{The spin gap of CaV$_4$O$_9$ revisited}
\vskip0.5truecm 
\author{M.~Mambrini and F.~Mila}

\address{Laboratoire de Physique 
Quantique, Universit\'e Paul Sabatier, 118 Route de Narbonne, 31062 
Toulouse Cedex, France. }\vskip0.5truecm
      
\maketitle

\begin{abstract}
\begin{center} 
\parbox{14cm}
{The large-plaquette scenario of the spin gap in CaV$_4$O$_9$ is investigated
on the basis of extensive exact diagonalizations. We confirm the existence
of a large-plaquette phase in a wide range of parameters, and we show that
the most recent neutron scattering data actually {\it require} an
intra-plaquette second neighbor exchange integral much larger than the inter-plaquette one,
thus justifying the perturbative calculation used in the interpretation of
the neutron scattering experiments.}
\end{center}
\end{abstract}
\vskip .1truein
 
\vskip2pc
]
CaV$_4$O$_9$ is the only known two-dimensional magnet presenting a gap
in the spin excitations. The low temperature dependence of the magnetic
susceptibility first measured by Taniguchi {\it et al.} \cite{taniguchi}
provided a value of $\Delta \approx 107$ K. Since then several scenarios
based on the Heisenberg model on the depleted lattice have been proposed to
explain both the presence and the value of $\Delta$. It appears that the
presence of frustration induced by a second neighbor antiferromagnetic
interaction is necessary and it has become clear that two types of first and
second neighbors exchange integrals are required to fit experiments correctly.
The Hamiltonian then reads,
\begin{eqnarray}
{\cal H} & = & J_{1} \sum_{{\langle i,j \rangle}_{p}} {\bf S}_i.{\bf S}_j
+J'_{1}\sum_{{\langle i,j \rangle}_{d}} {\bf S}_i.{\bf S}_j \nonumber \\
 & + & J_{2} \sum_{{\langle \langle i,j \rangle \rangle}_{p}} {\bf S}_i.{\bf S}_j
+J'_{2}\sum_{{\langle \langle i,j \rangle\rangle }_{d}} {\bf S}_i.{\bf S}_j
\label{hamiltonian}
\end{eqnarray}
where the sums run respectively over bonds within plaquettes (p) and
dimer between plaquettes (d) for first and second nearest neighbor.

The first explanation proposed to explain the presence of a gap was based on
a $J_{1}-J_{2}$ model on the depleted lattice with $J_{2}/J_{1}=1/2$ leading
to weakly coupled small plaquettes. It  must be abandoned for at
least two reasons: on one hand, recent estimates of the exchange
integrals in other members of the vanadates family suggest that the ratio $J_{2}/J_{1}$ is much larger
than $1/2$. On the other hand, the most recent data obtained with inelastic
neutron scattering seem to be inconsistent with the picture of weakly coupled
plaquettes built on the smallest squares of the depleted lattice. In particular, the dispersion curve of the lowest triplet state measured by Kodama {\it et
al.} \cite{kodama} turns out to be minimum at $(0,0)$ momentum. The authors
then performed fits of the curve assuming that one of the integral exchange is
large enough to justify a second order perturbation theory in the three other
parameters and  concluded that
the best set of exchange integral is the one corresponding to
$J_{1}$, $J'_{1}$, $J'_{2}$ small compared with $J_{2}$, namely
$J_{1}=J'_{1} \approx 67$ K, $J_{2} \approx 171$ K and $J'_{2} \approx 15$
K. These parameters are also consistent with the temperature dependence of
the susceptibility if one assumes that part of the sample is non-magnetic
\cite{takano}. In this picture the system consists of  a set of weakly coupled
$J_{2}$-plaquettes. Nevertheless, regarding the mechanism responsible
for the values of $J_{2}$ and $J'_{2}$ (super-exchange between
two second neighbor Vanadium atoms {\it via} an oxygen atom), the fact that
these integrals differ by one order of magnitude is {\it a priori}
surprising,  although it is not necessarily contradictory  since exchange
integrals are very difficult to evaluate directly (from {\it ab initio}
method in quantum chemistry for example). 

\begin{figure}
\begin{center}
\epsfig{file=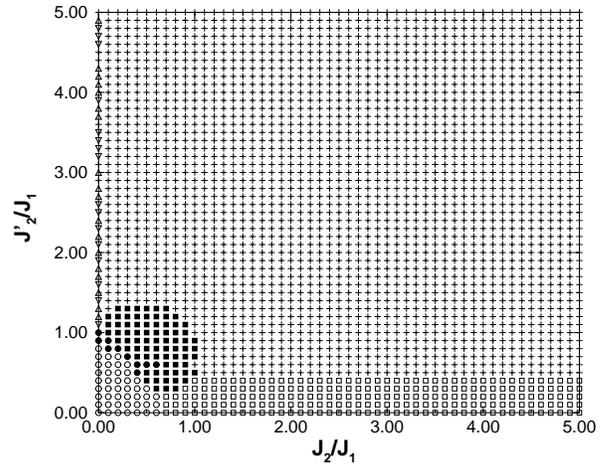,width=7.8cm}
\caption{Symmetry of the first excited state for $J'_{1}=J_{1}$.
Symmetry sectors are labeled by the momentum ${\bf k}=(k_x,k_y)$ and
the irreducible representation of the axial point group.  
$(0,0)\;A\;\;(S=0):(\bullet)$, $(0,0)\;B\;\;(S=0):(\blacksquare)$,
$(0,0)\;B\;\;(S=1):(\Box)$, $(0,0)\;x-iy\;\;(S=1):(\triangledown)$,
$(0,\pi)\;A\;\;(S=1):(+)$, $(\pi,\pi)\;A\;\;(S=1):(\circ)$,
$(\pi,\pi)\;x+iy\;\;(S=1):(\vartriangle)$. Note that the ground state
is in the $\bullet$ sector.}
\label{fig:map1}
\end{center}
\end{figure}

\begin{figure}
\begin{center}
\epsfig{file=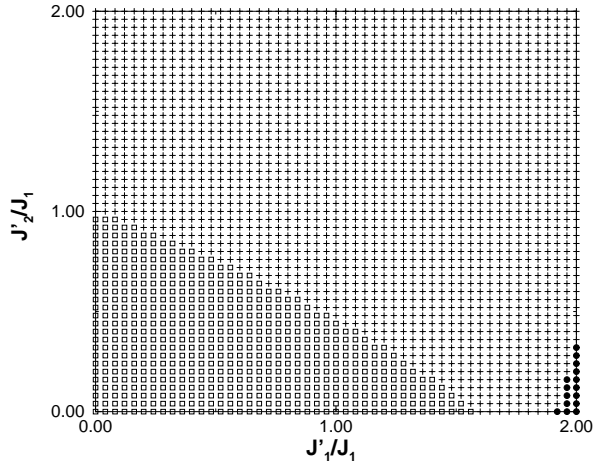,width=7.8cm}
\caption{Same as Fig. 1 for $J_{2}/J_{1}=2$.}
\label{fig:map2}
\end{center}
\end{figure}

To assess the validity of the hypotheses underlying Kodama {\it et al.}'s
analysis beyond perturbation 
theory, we have performed
exact diagonalizations on a 16 site cluster with periodic boundary conditions
without making {\it a priori} assumptions on the value of the exchange
integrals. First of all, we have determined the symmetry of the ground state and of
the first triplet excitation in the ($J_{2}/J_{1}$,$J'_{2}/J_{1}$) plane for
different values of $J'_{1}/J_{1}$ (see Fig. \ref{fig:map1} for
$J'_{1}=J_{1}$). Surprisingly enough, it turns out
that only rather small values of $J'_{2}/J_{1}$ are compatible with the
location of the minimum of the triplet band at ${\bf k} = (0,0)$. To check the robustness of this result with respect to variations of the
ratio $J'_{1}/J_{1}$,
we have done a similar calculation as a function of
($J'_{1}/J_{1}$,$J'_{2}/J_{1}$) for $J_{2}/J_{1}=2$ (see Fig. \ref{fig:map2}).
It shows that the restrictions are qualitatively the same, and that 
$J'_{2}/J_{1}$ must always be much smaller than $J_{2}/J_{1}$ in order to
reproduce the neutron scattering results.

\begin{figure}
\begin{center}
\epsfig{file=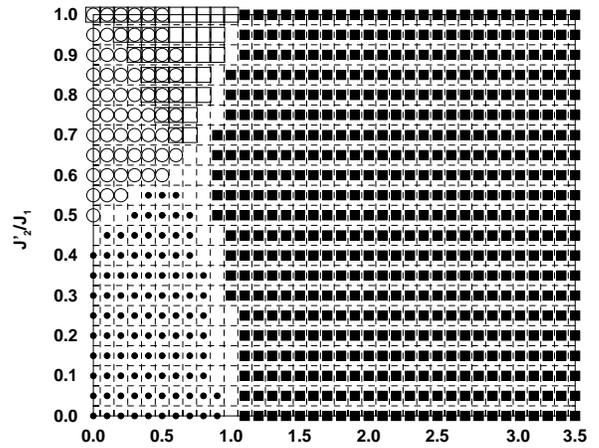,width=7.8cm}
\caption{``Phase diagram'' in the $(J_{2}/J_{1},J'_{2}/J_{1})$ plane. The
symbols denote points where $\langle {\bf S}_{i}.{\bf S}_{j}\rangle < -0.4$
for  $J_{1}$ ($\bullet$),$J'_{1}$ ($\circ$),$J_{2}$ ($\blacksquare$) and
$J'_{2}$ ($\Box$) bonds.}
\label{fig:correl}
\end{center}
\end{figure}

Next, we have checked whether the set of parameters used
in Ref. \cite{kodama} and \cite{takano} was consistent with
the picture of weakly coupled $J_{2}-$plaquettes. To answer this question we
have calculated the spin-spin correlation function 
$\langle {\bf S}_i.{\bf S}_j\rangle$ with $J'_{1}=J_{1}$ in the presumed range
of values of $J_{2}/J_{1}$ and $J'_{2}/J_{1}$. Fig. \ref{fig:correl} summarizes the results
and clearly delimits three different regions in which the system is
organized into small $J_{1}$-plaquettes, large $J_{2}$-plaquettes and
$J'_{1}-J'_{2}$ chains. We also checked that for $J_{2}/J_{1} \gtrsim 2$ the
correlation functions on all bonds except $J_{2}$ never exceed $0.1$ in absolute value, which shows
that the coupling between $J_{2}$-plaquettes is indeed very  small in that
range.

In conclusion, the results obtained by a systematic study of the energy
spectrum and of the spin-spin correlation functions of a 16 site cluster in a large range of
$J'_{1}/J_{1}$,$J_{2}/J_{1}$ and $J'_{2}/J_{1}$ ratios show that another
gapped phase made of weakly coupled $J_{2}-$plaquettes indeed exists when
$J_{2}/J_{1}$ is large enough. Besides, the experimental fact that the
minimum of the triplet band is  located at ${\bf k} = (0,0)$ implies that
$J'_{2}$ is much smaller than $J_{2}$. This suggests that the hypotheses
of Kodama {\it et al.} are indeed satisfied. Calculations on a 32 site
cluster to confirm the present results are in progress.

We thank IDRIS (Orsay) for allocation of CPU time on the C94 and C98
CRAY supercomputers.

\end{document}